# Design Patterns for Functional Strategic Programming

Ralf Lämmel
Vrije Universiteit
De Boelelaan 1081a
NL-1081 HV Amsterdam
The Netherlands
ralf@cs.vu.nl

Joost Visser
CWI
P.O. Box 94079
1090 GB Amsterdam
The Netherlands
Joost.Visser@cwi.nl

## ABSTRACT

In previous work, we introduced the fundamentals and a supporting combinator library for *strategic programming*. This an idiom for generic programming based on the notion of a *functional strategy*: a first-class generic function that cannot only be applied to terms of any type, but which also allows generic traversal into subterms and can be customized with type-specific behaviour.

This paper seeks to provide practicing functional programmers with pragmatic guidance in crafting their own strategic programs. We present the fundamentals and the support from a user's perspective, and we initiate a catalogue of *strategy design patterns*. These design patterns aim at consolidating strategic programming expertise in accessible form.

## 1. INTRODUCTION

Strategic programming is a novel generic programming idiom in which the notion of a *strategy* plays a crucial role [16]. In [17], we presented a realization of the strategic programming idiom in the functional programming paradigm, and we introduced the notion of a *functional strategy*. A functional strategy is a function with the following characteristics:

**generic** It can work on arguments of *any* type.

**specific** For specific types it can display customized behaviour.

**traversal** It can traverse into subterms.

**first-class** It can be named, passed as argument, etc.

The example in Figure 1 illustrates these characteristics. The function $increment$ is a functional strategy that increments all integers in a data structure by 1, regardless of the type of the data structure and of where the integers occur. This is demonstrated by its application to terms of type $[(Bool, Int)]$ and $Maybe\ (Int, ([Int], Int))$. The functions $topdown$, $adhoc$, and $identity$ are library combinators that will be discussed in more detail in Section 2.

As the example illustrates, the expressiveness of functional strategies goes beyond that of ordinary parametrically polymorphic and ad-hoc polymorphic functions. Note that $increment$ is essentially of type $\forall \alpha.\ \alpha \to \alpha$. The only parametrically polymorphic in-

$$increment = topdown\ (adhoc\ identity\ inc)$$
$$\textbf{where}\ inc :: Int \to Int$$
$$inc\ i = i + 1$$

$$increment\ [(True, 1)] \Longrightarrow [(True, 2)]$$
$$increment\ (Just\ (1, ([1], 1))) \Longrightarrow (Just\ (2, ([2], 2)))$$

**Figure 1: Example of a functional strategy.**

habitant of this type is the identity function, to which $increment$ is clearly not equivalent. Ad-hoc polymorphism is usually based on overloaded function declarations, but $increment$ is composed by customizing $identity$ with $inc$ using a function *combinator*, namely $adhoc$. The most direct way to type strategies involves the use of rank-2 types to point out that strategy combinators operate on generic functions. Alternatively, one can use first-class polymorphism or dynamic typing. In any case, additional effort is required to cope with traversal and type-specific customization. In Haskell, a range of different encodings are feasible [17, 14]. In the current paper, however, we will consider the types of strategies as *abstract*; we will not be bothered with their definition, only their use.

Strategic programming, i.e., program construction with strategies, constitutes a novel generic programming idiom with numerous benefits. It helps to attain separation of concerns, reusability, robustness, and conciseness when dealing with many-sorted data structures, such as documents and parse trees. The additional expressiveness of strategic programming has proven to pay off in application areas such as program transformation and analysis [4], reverse engineering [5], and grammar engineering [16].

Strafunski[1] is a Haskell-based bundle that supports generic programming with functional strategies. It contains an extensive library of reusable strategy combinators which can be composed, customized, and applied to construct application programs.

To effectively make use of the power that the idiom of strategic programming offers, more is needed than a combinator library. Deployment expertise must be gathered through practical experience, and consolidated in accessible form. In this paper, we take the perspective of the working functional programmer who wants to employ strategies. In Section 2, we review the fundamentals of strategies and we outline Strafunski's support for development of and with strategies. In Section 3 we present a catalogue of strategy design patterns. Each pattern is illustrated with code samples. Rather than choosing a trivial syntax to be processed by our sample strategies, or an arbitrary language, we have chosen samples dealing with the analysis or transformation of Haskell programs.[2]

---

[1] http://www.cs.vu.nl/Strafunski
[2] The corresponding system of datatypes is given in the appendix.

## 2. STRATEGIC PROGRAMMING

In this section, we discuss both the fundamentals of functional strategic programming, and Strafunski's support for it.

### 2.1 The essence of strategies

In the introduction, functional strategies were defined by enumerating their defining characteristics. They are functions that (i) work on arguments of any type, (ii) can display type-specific behaviour, (iii) can traverse into terms, and (iv) are first-class citizens. This abstract definition can be made more concrete by establishing a minimal set of basic strategy combinators that realizes these characteristics. Figure 2 shows such a set. Two strategy types are distinguished: $TP$ for type-preserving strategies (output type coincides with input type) and $TU$ for type-unifying strategies (output type is always $a$). Both types are parameterized with a monad $m$, such that monadic effects can be used in strategic programming.

The basic strategy combinators come in pairs: one for each strategy type as pointed out by the postfix "...TP" vs. "...TU". The $apply$ combinators justify our claim that strategies are generic functions (recall i). The $adhoc$ combinators support type-specific customization of a strategy (recall ii). If $f$ is a function on some type $T$, the strategy $adhoc\ s\ f$ will behave like $f$ when applied to a term of type $T$ and like $s$ on terms of any other type.

Out set contains the following nullary combinators. The *identity* combinator is a generic version of the monad member $return$ (which in turn is just the monadic identity function), i.e., $identity$ returns the input term. The *compute* combinator ignores the input term and always returns its argument. It is the generic counterpart of the $const$ function. The combinators $failTP$ and $failTU$ denote the always failing strategy in the sense of a *MonadPlus* with a member $mzero$ for failure. To instantiate the monad parameter $m$ of $TP$ and $TU$ with such a monad allows us to deal with partiality, recovery from failure, and backtracking.

The $seq$ and $let$ combinators perform their two argument strategies in sequence. For the $seq$ combinators, the first argument strategy is type-preserving, and its output is given to the second argument strategy as input. For the $let$ combinators, the first argument strategy is always type-unifying, and the second argument is a strategy parameterized with a value of the unifying type $a$. The result value of the first argument is used to instantiate the parameter of the second argument. The $choice$ combinators support recovery from failure of a strategy relying on the *mplus* member of the *MonadPlus* class. They attempt application of their first argument strategy, and if this fails, they apply their second argument strategy instead. In principle, the $choice$ combinators could also serve for a more general combination of alternatives in the sense of non-determinism or backtracking depending on the actual monad instance.

The $all$ and $one$ combinators are non-recursive generic traversal combinators (recall iii). In a sense, they push their argument strategy one level down into their input term. To be precise, $all$ applies its argument strategy to $all$ immediate subterms, while $one$ tries it left-to-right on each of the immediate subterms and stops after the first succeeds. Hence, the potentially failing strategy passed to the $one$ combinators involves *MonadPlus* to point out fitness of a child. The type-preserving variants of the traversal combinators preserve the outermost constructor of the input term. The type-unifying $allTU$ relies on the $mappend$ operator of a $Monoid$ to reduce the results of processing the subterms to a single result.

The $msubst$ combinators can be used to migrate from one monad to another. This is useful, for example, if we want to hide the fact that a certain strategy has the potential to fail while the overall strategy cannot due to recovery of failure. In this case we would migrate from the *Maybe* to the *Identity* monad.

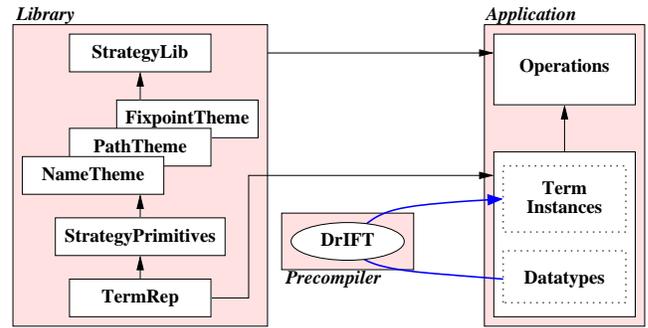

**Figure 3:** $Strafunski = Library + Precompiler$

Note that Strafunski's combinator style of generic programming indeed relies on the fact that strategies are first-class functions (recall iv): they can be named, passed as arguments, returned as results, and stored in data structures. Even traversal and type-specific customization are expressed via combinators.

### 2.2 Strafunski

Strafunski is a Haskell-based bundle that supports generic programming with functional strategies. Figure 3 provides an overview of the elements of Strafunski, and their relation to an application constructed with it. Strafunski consists of two components: a *library* and a *precompiler*, which we will discuss in turn.

*Library*

The library of Strafunski consists of a number of Haskell modules that address various aspects of strategic programming.

**StrategyLib** This is the top-level module of the library, provided for convenience. It allows the user to import the entire library with a single import statement.

**Themes** A series of modules is provided that covers a range of generic programming themes. For example, the *FixpointTheme* deals with iterative term transformation which terminates when some kind of fixpoint is found. The *TraversalTheme* defines various traversal schemes. The $NameTheme$ provides abstract algorithms for different kinds of name analysis, useful in language processing applications. In the $OverloadingTheme$, the basic strategy combinators of Figure 2 are overloaded to implement the intuition that the combinators usually come in pairs, one combinator for *TP* and another for *TU*. When defining new strategies, such overloading allows one to postpone commitment to a particular strategy type and, in a sense, to define two strategies at once. With each version of Strafunski, more themes are added and existing themes are elaborated. Excerpts of the current state of affairs are shown in Figure 4.

**StrategyPrimitives** This module provides basic strategy types, and a basic set of strategy combinators. Together they form an *abstract* datatype, whose internals are not exposed beyond the module. In fact, we have experimented with several implementations of the datatype that each have different characteristics with respect to performance, extensibility, and use of type features.

**TermRep** This module provides a generic term interface as a type class $Term$, as well as a universal representation of typed terms which is employed by this interface. The current implementation of the basic strategy combinators relies on *TermRep* to deal with dynamic typing and generic traversal while the module remains hidden for the rest of the library, and the user code.

| Strategy types | | | Description |
|---|---|---|---|
| **data** $TP\ m$ | = | $abstract$ | -- type-preserving |
| **data** $TU\ a\ m$ | = | $abstract$ | -- type-unifying |
| Basic combinators | | | |
| $applyTP$ | :: | $(Monad\ m, Term\ t) \Rightarrow TP\ m \to t \to m\ t$ | -- strategy application |
| $applyTU$ | :: | $(Monad\ m, Term\ t) \Rightarrow TU\ a\ m \to t \to m\ a$ | --   overloaded for a term type $t$ |
| $adhocTP$ | :: | $(Monad\ m, Term\ t) \Rightarrow TP\ m \to (t \to m\ t) \to TP\ m$ | -- strategy update |
| $adhocTU$ | :: | $(Monad\ m, Term\ t) \Rightarrow TU\ a\ m \to (t \to m\ a) \to TU\ a\ m$ | --   to add type-specific behaviour |
| $identity$ | :: | $Monad\ m \Rightarrow TP\ m$ | -- generic identity function |
| $compute$ | :: | $Monad\ m \Rightarrow m\ a \to TU\ a\ m$ | -- generic constant function |
| $failTP$ | :: | $MonadPlus\ m \Rightarrow TP\ m$ | -- always fail |
| $failTU$ | :: | $MonadPlus\ m \Rightarrow TU\ a\ m$ | --   using a monad for partiality |
| $seqTP$ | :: | $Monad\ m \Rightarrow TP\ m \to TP\ m \to TP\ m$ | -- perform strategies in sequence |
| $seqTU$ | :: | $Monad\ m \Rightarrow TP\ m \to TU\ a\ m \to TU\ a\ m$ | --   when the first strategy is type-unifying |
| $letTP$ | :: | $Monad\ m \Rightarrow TU\ a\ m \to (a \to TP\ m) \to TP\ m$ | --   its result value is passed as argument |
| $letTU$ | :: | $Monad\ m \Rightarrow TU\ a\ m \to (a \to TU\ b\ m) \to TU\ b\ m$ | --   to the second strategy |
| $choiceTP$ | :: | $MonadPlus\ m \Rightarrow TP\ m \to TP\ m \to TP\ m$ | -- attempt alternative strategies |
| $choiceTU$ | :: | $MonadPlus\ m \Rightarrow TU\ a\ m \to TU\ a\ m \to TU\ a\ m$ | --   using the "+" of an extended monad, e.g., $Maybe$ |
| $allTP$ | :: | $Monad\ m \Rightarrow TP\ m \to TP\ m$ | -- apply argument to all immediate subterms |
| $allTU$ | :: | $(Monad\ m, Monoid\ a) \Rightarrow TU\ a\ m \to TU\ a\ m$ | --   for $TU$, results are reduced with a monoid's "+" |
| $oneTP$ | :: | $MonadPlus\ m \Rightarrow TP\ m \to TP\ m$ | -- apply argument to one kid |
| $oneTU$ | :: | $MonadPlus\ m \Rightarrow TU\ a\ m \to TU\ a\ m$ | --   try kids from left to right, fail if none succeed |
| $msubstTP$ | :: | $(Monad\ m, Monad\ n) \Rightarrow (\forall\ t\ .\ m\ t \to n\ t) \to TP\ m \to TP\ n$ | -- substitute one monad by another |
| $msubstTU$ | :: | $(Monad\ m, Monad\ n) \Rightarrow (m\ a \to n\ a) \to TU\ a\ m \to TU\ a\ n$ | --   using a helper function |

**Figure 2: Strategy types and basic strategy combinators.**

| Fixpoint theme | | | Description |
|---|---|---|---|
| $repeatTP\ s$ | = | $tryTP\ (seqTP\ s\ (repeatTP\ s))$ | -- keep applying $s$ until it fails |
| $outermost\ s$ | = | $repeatTP\ (oncetd\ s)$ | -- outermost evaluation strategy |
| $innermost\ s$ | = | $repeatTP\ (oncebu\ s)$ | -- innermost evaluation strategy |
| Traversal theme | | | |
| $topdown\ s$ | = | $s\ `seqTP`\ (allTP\ (topdown\ s))$ | -- apply $s$ in topdown fashion to all nodes |
| $bottomup\ s$ | = | $(allTP\ (bottomup\ s))\ `seqTP`\ s$ | -- idem, in bottom up fashion |
| $stoptd\ s$ | = | $s\ `choiceTP`\ (allTP\ (stoptd\ s))$ | -- cutoff traversal below nodes where $s$ succeeds |
| $oncetd\ s$ | = | $s\ `choiceTP`\ (oneTP\ (oncetd\ s))$ | -- terminate traversal at first node where $s$ succeeds |
| $oncebu\ s$ | = | $(oneTP\ (oncebu\ s))\ `choiceTP`\ s$ | -- idem, bottom up |
| $crush\ s$ | = | $comb\ mappend\ s\ (allTU\ (crush\ s))$ | -- combine results of applying $s$ to all nodes with + of monoid |
| $select\ s$ | = | $s\ `choiceTU`\ (oneTU\ (select\ s))$ | -- return result of first succesful application of $s$ (topdown) |
| Control & data-flow theme | | | |
| $build\ a$ | = | $compute\ (return\ a)$ | -- variant of $compute$ with non-monadic domain |
| $tryTP\ s$ | = | $s\ `choiceTP`\ identity$ | -- recover from failure |
| $comb\ o\ s\ s'$ | = | $s\ `letTU`\ \lambda a \to s'\ `letTU`\ \lambda b \to build\ (o\ a\ b)$ | -- combine the result of two strategies with binary operator $o$ |
| $before\ s\ f$ | = | $s\ `letTU`\ compute \circ f$ | -- apply type-unifying strategy $s$ and then $f$ to its result |
| Container theme | | | |
| $modifyTU\ f\ t$ | = | $adhocTU\ f \circ modify\ (applyTU\ f)\ t$ | -- generic pointwise function update |
| | | **where** $modify\ f\ x\ y$ | -- defined by lifting |
| | | $= \lambda x' \to$ **if** $x \equiv x'$ **then** $y$ **else** $f\ x'$ | -- non-generic counterpart |

**Figure 4: Excerpts from Strafunski's theme modules.**

*Precompiler*

To use the Strafunski library in an application, instances of the *Term* class must be provided for the datatypes of the application. This can be done manually, but Strafunski provides a precompiler to automate the process. This is possible because these *Term* instances follow a very simple scheme for all algebraic datatypes. Currently, the precompiler is implemented as an extension of the DrIFT tool (formerly known as Derive [22]). If the following directive is added to a Haskell source file:

```
{-! global: Term !-}
```

the precompiler will generate and insert *Term* instances for all datatypes in the file. For the sample code in the present paper, we precompiled the abstract syntax of Haskell 98 (see the appendix) to enable traversal over Haskell parse trees.

Thus, functional strategic programming with Strafunski proceeds along the following steps:

1. Apply the precompiler to the system of datatypes that represent the terms on which to operate.

2. Import the precompiled datatypes and $StrategyLib$ into an application module.

3. Select, combine, and specialize appropriate strategy combinators from Strafunski's library, and apply the resulting strategies to the terms that need to be processed.

Clearly, the last of these steps deserves elaboration.

| NAME | CATEGORY |
|---|---|
| **Aka\*** | |
| Alternative names. | |
| **Intent** | |
| Short statement of the pattern's purpose. | |
| **Motivation** | |
| Description of a particular design problem and a brief indication of how the pattern can be used to solve it. | |
| **Applicability\*** | |
| Preconditions for using the pattern. | |
| **Schema** | |
| A schematic code fragment that indicates the participants in the pattern and their relationships. | |
| **Description** | |
| Explanation of the schema that details the responsibilities of all participants and describes how they collaborate to carry them out. | |
| **Sample code** | |
| Actual working Haskell code in which the pattern is used. | |
| **Consequences\*** | |
| Description of the results and trade-offs of applying the pattern. | |
| **Related patterns\*** | |
| Similarities, differences, and connections to other patterns. | |

**Figure 5: Format of each design pattern description.**

## 3. DESIGN PATTERNS

The novelty of the strategic programming idiom implies that few are experienced and well-versed in it. Though Strafunski's library provides an extensive array of predefined strategy combinators, deployment of these combinators for program construction is an acquired skill, as is any (functional) programming style. In this section, we attempt to convey our (limited) deployment expertise in a set of *design patterns*.

### *What is a design pattern?*

The notion of a *design pattern* is well-established in object-oriented programming. In the first pattern catalogue, design patterns are defined as *"descriptions of communicating objects and classes that are customized to solve a general design problem in a particular context"* [7, page 3]. Each pattern systematically names, motivates, and explains a common design structure that addresses a certain group of recurring program construction problems. After the initial 23 patterns of the first catalogue, numerous further patterns have been described in roughly the same style and format.

We contend that design patterns can be an effective means of consolidating and communicating program construction expertise for functional programming just as they have proven to do in object-oriented programming. Of course, a few modifications are in order to accommodate the characteristics of functional programming and the space limitations of this publication. Figure 5 shows the format we have chosen to describe each pattern. This format is very similar to the one known from object-oriented literature. We have chosen to make particular items optional (indicated by an asterisk). Also, the diagrams of object-oriented class structures seem to have no obvious functional counterparts, so we provide schematic Haskell code fragments instead. Finally, it seems that the conciseness of functional programming with respect to object-oriented programming carries over to our strategic design patterns.

| Pattern | Concerns |
|---|---|
| REWRITE STEP | non-generic computation step |
| GENERIC REWRITE STEP | generic computation step |
| TRAVERSAL | traversal behaviour *vs.* computation step |
| KEYHOLE OPERATION | strategic behaviour *vs.* strategy-free interface |
| SUCCESS BY FAILURE | traversal control |
| CIRCUITRY | control- and data-flow *vs.* computation steps |
| ROLE PLAY | analyses *vs.* guards *vs.* side-effects *vs.* transformations |
| TRAVERSAL SCHEME | purely generic traversal behaviour |
| PROPAGATION | environment passing |
| LOCAL EFFECT | effectful behaviour *vs.* effect-free interface |
| GENERIC CONTAINER | heterogeneous datatypes |
| TYPE ARGUMENT | type-specific behaviour *vs.* specifically typed values |
| META SCHEME | skeleton of a traversal scheme |

**Figure 6: Concerns that each pattern isolates or separates.**

### *Organizing the catalogue*

Each design pattern is aimed at solving only a single design problem. Clearly, in actual applications design problems never come alone, and combinations of patterns must be used. The selection of appropriate design patterns for a given set of design problems can be guided by categorizing the patterns according to various criteria. We briefly discuss three such criteria.

Firstly, we have divided our design patterns into two main groups:

**Basic** Patterns in this group address design problems encountered by any strategic programmer. Required reading.

**Advanced** Patterns in this group address less frequent design problems. Read these when you are ready to bring your strategic programming skill to a higher level.

Secondly, each design pattern can be characterized by the kind of isolation and separation of concerns that can be accomplished with them. Figure 6 provides an overview. To give an example, the design pattern TRAVERSAL SCHEME isolates the concern of purely generic traversal behaviour in the sense that all type-specific behaviour will be supplied by the instantiation of a traversal scheme.

Thirdly, we can categorize by used means of parameterization. We can classify the parameters of functional strategies with respect to several dimensions: (i) whether it is monomorphic or polymorphic, (ii) whether explicit quantification inside the strategy type is used, or implicit quantification at the top level, (iii) whether it is intended to contribute type-specific behaviour, (iv) whether it is intended to involve traversal behaviour, and (v) what its order and arity are. To give an example, let us sketch how a traversal over some application-specific data is organized according to the design pattern TRAVERSAL. We first select for instance a traversal scheme with one strategy argument. So it is a unary first-order strategy combinator (see v). The strategy argument is polymorphic (see i), and it is explicitly quantified (see ii). The argument is supposed to contribute type-specific behaviour (see iii), but traversal behaviour is not mandatory (see iv). Parameterization is the prime abstraction mechanism offered by functional programming. The strategy design patterns help a generic programmer to put this abstraction mechanism to work.

## Rewrite Step     Basic

**Intent**
Capture a single type-specific computation step.

**Motivation**
Generic programming involves type-specific and generic functionality. By capturing type-specific computations and assigning a name to them, they can easily be reused in different contexts. A rewrite step is such a reusable piece of type-specific functionality.

**Schema**

$$step :: T \to T'$$
$$step\ pat = rhs$$
$$step\ v = ...$$

**Description**
Model a rewrite step with a unary function *step* on a specific term type $T$. The result type $T'$ may or may not coincide with $T$, and may or may not be monadic. Define the function with equations that pattern-match on the argument. If the pattern-match cases are not exhaustive, then the function needs to be complemented by a catch-all case. You can use the *Maybe* type constructor to indicate when the step fails to fire. Alternatively, the catch-all equation can return the input term itself, or a distinguished value, such as the empty list.

**Sample code**
Return the type constructor name from a type expression

$$refTypes :: HsType \to [HsName]$$
$$refTypes\ (HsTyCon\ (UnQual\ n)) = [n]$$
$$refTypes\ \_ = [\,]$$

Return the type constructor name from a type declaration

$$decTypes :: HsDecl \to [HsName]$$
$$decTypes\ (HsTypeDecl\ \_\ n\ \_\ \_) = [n]$$
$$decTypes\ (HsDataDecl\ \_\ \_\ n\ \_\ \_\ \_) = [n]$$
$$decTypes\ (HsNewTypeDecl\ \_\ \_\ n\ \_\ \_\ \_) = [n]$$
$$decTypes\ \_ = [\,]$$

The two above rewrite steps deal with name analysis for Haskell programs. They work on the abstract syntax of type expressions and declarations, respectively. The first step retrieves the name of a type constructor referred to by the type expression, if any. The second step retrieves the name of a type constructor declared in the given declaration, if any. We use lists of names as result type so that we are able to deal with cases where there is one type name, no type name, and potentially even several type names. In both cases, we need a catch-all case because obviously not all syntactical patterns are covered by the pattern-match cases. The catch-all cases simply return the empty list.

**Consequences**
By capturing relatively small pieces of type-specific behaviour in separate rewrite rules, this behaviour can be used as building blocks for larger strategic programs.

**Related Patterns**
The construction of rewrite steps is a prerequisite for creating Generic Rewrite Steps, but they can also be passed as actual parameters to Keyhole Operations.

## Generic Rewrite Step     Basic

**Intent**
Lift type-specific rewrite steps to the strategy level, making them applicable to terms of all types.

**Motivation**
Each individual rewrite step captures a computation that deals with data of a single type. At some point in the synthesis of generic programs, type-specific rewrite steps need to be made generic. This involves the composition of possibly several type-specific rewrite steps (for different types) and the indication of a generic default for all the types that are not covered by the type-specific computations. This entire composition is called a generic rewrite step.

**Applicability**
The non-generic rewrite steps composed into a generic one must be specific for *different* types. To compose rewrite steps that are specific for the same type, use the *choice* combinator, following Success By Failure.

**Schema**

$$poly = def\ `adhoc`\ (\ldots\ s_1\ \ldots)\ \ldots\ `adhoc`\ (\ldots\ s_n\ \ldots)$$

**Description**
To compose rewrite steps $s_1, \ldots, s_n$ for different types into a strategy for *any* type, use (repeated application of) the *adhoc* combinators. Start from a default strategy *def* to deal with all types not covered by $s_1, \ldots, s_n$. Typical defaults are the strategies *fail*, *identity*, or *build*. Since strategies are always monadic entities in Strafunski, you must make the specific rewrite steps monadic, if they are not yet by themselves. The *identity* monad is the most basic choice.

**Sample code**
Return constructor names that are declared or referred to

$$anyTypes :: TU\ [HsName]\ Identity$$
$$anyTypes = build\ [\,]$$
$$\phantom{anyTypes = }`adhocTU`\ (return \circ decTypes)$$
$$\phantom{anyTypes = }`adhocTU`\ (return \circ refTypes)$$

In this sample we reuse the type-specific rewrite steps *decTypes* and *refTypes* that illustrate the Rewrite Step pattern. These steps are specific to the types *HsDecl* and *HsType*. We combine them into a single generic rewrite step with $build\ [\,]$ as default. The composed function identifies type constructor names in a given term, both in declaration and use sites. The chosen generic default specifies that the empty list should be returned when faced with terms of any other type than *HsDecl* or *HsType*. The non-generic rewrite steps are made monadic by composing them with *return*. We have opted for the trivial *Identity* monad.

**Consequences**
By making a rewrite step generic, it can be applied to terms of any type, and it becomes amenable to specialization with further type-specific behaviour.

**Related Patterns**
Lifting rewrite steps to the strategy level, i.e., turning them into generic rewrite steps is a prerequisite for passing them as arguments to a Traversal Scheme, and for using them as building blocks in Circuitry.

## TRAVERSAL                                                           BASIC

**Intent**
Instantiate a traversal scheme with generic rewrite steps.

**Motivation**
Traversal is at the heart of strategic programming. Many schemes of traversal are readily available in Strafunski's combinator library. You can construct a traversal by passing your own rewrite steps to an appropriate predefined traversal scheme.

**Schema**

$$instantiation = scheme \; ap_1 \; \ldots \; ap_n$$
$$\textbf{where} \; ap_1 = \ldots \text{`adhoc`} \ldots$$
$$\ldots$$
$$ap_n = \ldots \text{`adhoc`} \ldots$$

**Description**
To select an appropriate traversal scheme from the library you must first decide whether you need a type-unifying one (for analysis) or a type-preserving one (for transformation). Further, you must decide on the desired order of traversal (e.g., top-down or bottom-up), whether the traversal should be cut-off below certain nodes (stop conditions), how to combine intermediate results, and more. These decision will usually lead to the identification of a library scheme. Then, you have to identify the rewrite steps required to solve the problem. These rewrite steps are usually generic, and they serve as the actual parameters $ap_1, \ldots, ap_n$ that *instantiate* the traversal *scheme*.

**Sample code**
Collect all type constructor names from a given term

$$allTypes :: TU \; [HsName] \; Identity$$
$$allTypes = crush \; anyTypes$$

Using the predefined combinator *crush* with type:

$$crush :: (Monad \; m, Monoid \; a) \Rightarrow TU \; a \; m \to TU \; a \; m$$

The strategy *allTypes* uses the generic rewrite step *anyTypes*, that illustrates the GENERIC REWRITE STEP pattern, to collect all possible declaring and referring occurrences of type constructor names. For this purpose we selected the traversal scheme *crush*, which performs deep reduction in top-down order over the entire input term (no cut-off). The intermediate results are combined via a *Monoid*'s binary operator. Since we use lists as monoids, this binary operator will be resolved to the append operator "++". The resulting traversal can be applied to terms of any type. An example of a type-preserving traversal is provided by the *increment* strategy of Figure 1. Note that, to adhere to Strafunski's naming conventions in this figure, *adhocTP* should replace *adhoc*, and *applyTP* should be prefixed to the application examples.

**Related Patterns**
If you discover that the particular traversal scheme you need is not present in Strafunski's library, then you may consult TRAVERSAL SCHEME to find out how to roll your own. Traversals can be disguised by a KEYHOLE OPERATION, and can be used to fulfill various roles in a ROLE PLAY.

## KEYHOLE OPERATION                                                   BASIC

**Aka**
Wrapper Worker

**Intent**
Do not expose strategies to the top level.

**Motivation**
In the Strafunski-style, strategies are generic functions, subject to function application via *applyTP* and *applyTU*. When they serve as argument or result, this can be observed via the types *TP* and *TU*. If you want to use strategies without exposing them to the outside, you can use a keyhole operation. On the inside, you can dispose of the full power of strategies, while on the outside, all you see is a plain function without any trace of *TP* or *TU*.

**Schema**

$$wrapper \; fp_1 \; \ldots \; fp_n = \ldots \; apply \; worker \; \ldots$$
$$\textbf{where} \; worker = \ldots \; (\ldots \text{`adhoc`} \; fp_1) \; \ldots$$
$$\ldots$$
$$\ldots \; (\ldots \text{`adhoc`} \; fp_n) \; \ldots$$

**Description**
Divide the functionality of your algorithm over a top-level *wrapper* function which directly operates on terms, and a nested *worker* strategy. Use *adhoc* when specific argument strategies are used in the definition of the worker. Use *apply* to define the specific wrapper in terms of the generic worker.

**Sample code**
Check whether a Haskell type constructor is fresh

$$isFreshType :: HsName \to HsModule \to Bool$$
$$isFreshType \; n = runIdentity \circ applyTU \; worker$$
$$\textbf{where}$$
$$worker = allTypes \; \text{`before`} \; isNotElem$$
$$isNotElem = return \circ \neg \circ elem \; n$$

General focus selection

$$selectFocus :: (MonadPlus \; m, Term \; f, Term \; t)$$
$$\Rightarrow (f \to m \; f) \to t \to m \; f$$
$$selectFocus \; getFocus = applyTU \; worker$$
$$\textbf{where}$$
$$worker = select \; (adhocTU \; failTU \; getFocus)$$

The operation *isFreshType* implements a predicate to test if a certain type constructor name $n$ is fresh (i.e., not yet used) in a given Haskell module. Note that this is a completely monomorphic function. The wrapped worker is basically the traversal *allTypes* that illustrates the TRAVERSAL pattern but postfixed by a membership test *isNotElem*. After wrapping the worker we further postprocess the result with *runIdentity* to get out of the trivial *Identity* monad. The second example above deals with selection of terms from a focus where the helper *getFocus* for focus detection is passed to a keyhole operation. Internally, the traversal scheme *select* is used.

**Consequences**
With a keyhole operation you fit a non-generic interface on generic functionality. This means you can apply the wrapper function with ordinary function application, instead of using *apply*. On the other hand, if you want to pass the generic functionality to a traversal scheme, or update it with *adhoc*, you will have to go around the wrapper and use the worker directly.

## SUCCESS BY FAILURE — BASIC

**Intent**
Use a potentially failing computation to control traversal.

**Motivation**
To correctly implement certain traversals, their constituent rewrite steps should only be performed under certain conditions. For instance, a particular rewrite step should trigger only if another does not, or vice versa. To model success and failure of rewrite steps and strategies, you can use the *Maybe* monad or a backtracking monad. Generic failure is captured by the *fail* combinators, and the *choice* combinators allow you to recover from failure.

**Schema**

$$partial = gstep\ `choice`\ \ldots$$
$$\textbf{where}$$
$$gstep = fail\ `adhoc`\ step$$
$$step\ pat = return\ \ldots$$
$$step\ v = mzero$$

**Description**
A *partial* strategy, i.e., one that potentially fails, is typically constructed from type-specific rewrite *step*s that use the *mzero* of a *MonadPlus* to encode failure. When lifting such a partial *step* to the strategy level with *adhoc*, the generic *fail* combinator is used as default strategy. Finally, the *choice* combinator is used to combine potentially failing generic rewrite steps.

**Sample code**
The library scheme for selection

$$select :: MonadPlus\ m \Rightarrow TU\ a\ m \to TU\ a\ m$$
$$select\ s = s\ `choiceTU`\ (oneTU\ (select\ s))$$

Identify different kinds of type constructor names

$$decCon = choice\ TU\ (failTU\ `adhocTU`\ typeCon)$$
$$\qquad\qquad\qquad (failTU\ `adhocTU`\ dataCon)$$
$$\textbf{where}$$
$$typeCon\ (HsTypeDecl\ \_\ n\ \_\ \_) = return\ n$$
$$typeCon\ \_ = mzero$$
$$dataCon\ (HsDataDecl\ \_\ \_\ n\ \_\ \_\ \_) = return\ n$$
$$dataCon\ \_ = mzero$$

A prime example of a partial strategy combinator is *select*. Its argument strategy is meant for the identification of selectable entities. This process must be necessarily partial. A *choice* is used in the definition of *select* because selection can recover from failure of identification for a given node by recursing into the children. If the identification strategy fails at all levels, selection will altogether fail.

The second sample illustrates the use of potentially failing strategies to merge rewrite steps that are specific for the same type. The partial rewrite steps *typeCon* and *dataCon* are both specific for type *HsDecl*. The *decCon* strategy reverts to the second if the first fails.

**Related Patterns**
If you want to prevent the monadic effects of partiality or non-determinism to invade parts of your code that do not rely on them, you may want to use LOCAL EFFECT.

## CIRCUITRY — BASIC

**Intent**
Use composition and recursion to assemble strategies into a composite traversal with appropriate control and data flow.

**Motivation**
When composing traversals, one should take care to sequence the ingredient steps in the right order, to pass data to the steps that need them, and to traverse the appropriate parts of the input term. By connecting your steps with appropriate combinators and recursive calls you can wire up the control and data flow between them.

**Schema**

$$strategy = \ldots\ strategy\ \ldots$$
$$\qquad\qquad `co_1`$$
$$\qquad\qquad \ldots\ strategy\ \ldots$$
$$\qquad\qquad \ldots$$
$$\qquad\qquad `co_n`$$
$$\qquad\qquad \ldots\ strategy\ \ldots$$

**Description**
Typical choices for the composition operators $co_i$ are the *seq*, *let*, and *choice* combinators. The *seq* operators are used to prefix a strategy with a type-preserving strategy. The *let* operators are used to compute a value via a type-unifying strategy and to pass it on. The *choice* operators are meant for branching control-flow. All the composed strategies potentially include recursive references to *strategy*.

**Sample code**
Compute free variables in a given Haskell fragment

$$freeHsVars :: TU\ [HsName]\ Identity$$
$$freeHsVars = refHsVars\ `letTU`\ \lambda refs \to$$
$$\qquad\qquad decHsVars\ `letTU`\ \lambda decs \to$$
$$\qquad\qquad allTU\ freeHsVars\ `letTU`\ \lambda frees \to$$
$$\qquad\qquad build\ (union\ frees\ refs\ \backslash\backslash\ decs)$$
$$\textbf{where}$$
$$refHsVars, decHsVars :: TU\ [HsName]\ Identity$$
$$refHsVars = adhocTU\ (build\ [])\ (return \circ step)$$
$$\qquad\textbf{where}$$
$$\qquad step\ (HsVar\ (UnQual\ n)) = [n]$$
$$\qquad step\ \_ = [\,]$$
$$decHsVars = \ldots$$

The above strategy performs free variable analysis on arbitrary Haskell program fragments. Free variables are obtained by subtracting (cf. "$\backslash\backslash$") the locally declared variables *decs* from the *union* of the locally referenced variables *refs* and the free variables *frees* from the subterms. We use two generic rewrite steps *refHsVars* and *decHsVars* for the identification of declaring and referring occurrences of Haskell variables. The strategy is recursively defined to descend into terms via *allTU*. The *letTU* combinator is used to connect all the type-unifying computations.

**Related Patterns**
The SUCCESS BY FAILURE pattern shows how partiality of strategies can be modeled, and how it can be used to realize branches in the control and data-flow between your strategic components.

## Role Play — Advanced

**Intent**
Define a transformation as a pipeline of steps with designated roles.

**Motivation**
A transformation can usually be decomposed into separate steps with limited responsibilities, such as analyses, guards, side effects, atomic transformations, and others. When each step has its own sharply delimited role to play, it becomes easier to construct, understand, and modify the transformation. The individual steps can be formed into a complete transformation pipeline with appropriately selected composition operators.

**Schema**

$$transformation = role_1 \; `co_1` \; role_2 \; \ldots \; `co_n` \; role_n$$

**Description**
Decompose the transformation task you need to implement into basic roles. An *analysis* is type-unifying, and does not modify its input term. A *guard* checks whether a particular condition is satisfied by its input term. It is typically implemented as a Boolean expression wrapped by the *guard* function, or a strategy of type $TU\;()\;m$, where the monad *m* supports partiality. Side effects are realized by access to an extended monad interface for a state. Atomic *transformation steps* are type-preserving.
There are two kinds of pipelines. Depending on whether you need the pipeline itself to be a strategy, you may either compose strategies, or keyhole operations and other monadic functions. The composition operators $co_i$ are *let* and *seq* combinators when composing strategies. The monadic bind operator ">>=", or do-notation are used when composing keyhole operations.

**Sample code**
Replace a focussed type expression by a type synonym

```
toAlias :: HsName → HsModule → Maybe HsModule
toAlias n m =
   do t ← selectTypeFocus m
      t' ← getAlias n m
      guard (t ≡ t')
      replaceTypeFocus n m
```

The *toAlias* pipeline implements a simple refactoring for Haskell datatypes. Assuming that a focus has been placed on some type expression *t*, we want to replace *t* by a type synonym (or alias) named *n*. A pre-condition for this replacement is that *n* is defined as *t* in the given Haskell module *m*. The transformation is implemented as a sequence of keyhole operations and a simple guard. For brevity, we do not show the definitions of the keyhole operations. Firstly, we look up the type expression *t* from the focus via *selectTypeFocus*. This is an analysis. Secondly, we look up the right-hand side expression *t'* from the declaration for *n* via *getAlias*. This is again an analysis. Then, we place a guard to enforce that the focused type expression *t* actually coincides with *t'*. Finally, we perform the actual transformation that replaces the focussed type expression by a reference to *n* via *replaceTypeFocus*.

**Related Patterns**
The CIRCUITRY pattern explains how to wire the data and control flow between the individual steps of a pipeline.

## Traversal Scheme — Advanced

**Aka**
Abstract Algorithm, Recursion Scheme.

**Intent**
Capture traversal control in a fully generic, reusable strategy combinator, which abstracts over any type-specific operations.

**Motivation**
The traversal behaviour of many traversals can be captured in a reusable traversal *scheme*. To ensure its reusability, type-specific computations should not be hard-wired into it, but should rather be supplied via appropriate parameters. Thus, a traversal scheme captures generic traversal behaviour in an abstract algorithm.

**Schema**

$$scheme\;fp_1\;\ldots\;fp_n = \\ \ldots\;fp_1\;\ldots\;fp_n\;\ldots\;(scheme\;fp_1\;\ldots\;fp_n)\;\ldots$$

**Description**
Divide your algorithm into a fully generic *scheme* with formal parameters for type-specific computations. These parameters are either of strategy types, or they are place holders for monomorphic functions. The scheme itself should not make use of *adhoc* combinators. Rather, the actual parameters that are supplied when the scheme is instantiated should implement type-specific behaviour via *adhoc* combinators.

**Sample code**
Generic free name analysis

```
freeNames :: Eq n
             ⇒ TU [n] Identity
             → TU [n] Identity
             → TU [n] Identity
freeNames refNames decNames = fnames
   where
      fnames = refNames `letTU` λrefs →
               decNames `letTU` λdecs →
               allTU fnames `letTU` λfrees →
               build (union frees refs \\ decs)
```

Instantiations

```
freeHsVars = freeNames refHsVars decHsVars
freeHsTVars = freeNames refHsTVars decHsTVars
freeJaVars = freeNames refJaVars decJaVars
```

The free names (e.g., variables) in a given program fragment can be collected by a strategy which looks up the names from all the relevant patterns dealing with names in the given language. The sample code for the CIRCUITRY pattern defines a Haskell-specific free variable analysis. The above *freeNames* combinator implements a generic scheme for free name analysis by separating out the type-specific ingredients of the traversal. By supplying appropriate actual parameters for recognition of referred variables and declared variables, we can obtain different concrete name analysis algorithms, e.g., for free Haskell variables, free Haskell *type* variables, or free Java variables.
Examples of simpler traversal schemes are the predefined combinators of Strafunski's *TraversalTheme* (see Figure 4).

**Related Patterns**
Instantiation of a traversal scheme to synthesize an actual traversal is described in the TRAVERSAL pattern.

## LOCAL EFFECT     ADVANCED

**Intent**
Do not expose monadic effects beyond where they are needed.

**Motivation**
Effects such as partiality, non-determinism, and state can be used in strategic programming by employing appropriate (stacked) monads. Often, such effects are only needed locally. With a local effect you can prevent locally needed monads to pollute the rest of your program.

**Schema**

```
effectful :: TP MEffect
effectful = ... effect ...

effectless :: TP M
effectless = msubstTP m2m effectful
  where m2m :: MEffect → M
        m2m = return ∘ runME
```

**Description**
Implement the functionality that requires an effect in a combinator that exposes the corresponding monad *MEffect*. Call this *effectful* combinator from a second combinator that exposes a different monad *M*, without the effect. Use the *msubst* combinator to substitute one monad by the other, using a function *m2m* that runs the effectful computation and returns its value inside the monad without effect. Instead of using unrelated monads, you can construct *MEffect* by applying a monad transformer to *M*. The function *m2m* should then 'unlift' the transformed monad to recover the original monad.

**Sample code**
Localize a state transformer

```
localStT :: Monad m ⇒ s → TP (StateT s m) → TP m
localStT s
  = msubstTP unlift
    where unlift tm = evalStateT tm s
```

Replace all strings with a deBruijn index

```
deBruijn :: Monad m ⇒ TP m
deBruijn
  = localStT "1" (topdown poly)
    where poly = adhocTP identity mono
          mono = λ_ → do n ← get
                         put (n ++ "'")
                         return n
```

The *deBruijn* strategy replaces all Strings in a given input term by unique identifiers, starting with `"1"`, and then adding a prime at each step. Internally, a state monad transformer is used to keep track of the most recently generated identifier. Externally, i.e., looking at the type of *deBruijn*, there is no trace of this state monad. This is accomplished with the *localStT* combinator, which converts a strategy that employs a monad with state transformer into a strategy on the same monad, without the transformer. The parameter *s* represents the initial state. The conversion is accomplished by calling *msubstTP* with a function *unlift* that evaluates the state transformer initialized with *s*.

**Consequences**
Localizing a monadic effect can improve not only readability of your code. The performance of your program may benefit as well.

## PROPAGATION     ADVANCED

**Aka**
Hand Me Down

**Intent**
Propagate data downwards into the traversed tree.

**Motivation**
What you do with lower nodes in the tree might be dependent on information collected or constructed at higher nodes. With Propagation, such information is handed down via a parameter of the recursive call of a traversal.

**Schema**

```
traversal :: e                    -- initial data
          → (e → TU e m)          -- data modifier
          → (e → S)               -- node action
          → S                     -- traversal
traversal e modify action
        = ...
            action e
            ...
            modify e 'letS' λe' →
            traversal e' action modify
            ...
```

**Description**
To add data propagation behaviour to a traversal, you should first parameterize the node action(s) of your traversal with the type of this data. Furthermore, you should add two parameters to your *traversal*. First, initial data to start the traversal with. Second, a function to modify the data at each step downward during traversal. This function takes current data and current node as input, and computes new data. At each node, three things happen. Firstly, the node *action* is applied, using the current data *e*. Secondly, the current data is modified. Thirdly, the new data is used in a recursive call of the complete *traversal*.

**Sample code**
A propagating version of the traversal scheme *select*

```
selectenv :: MonadPlus m
          ⇒ e → (e → TU e m) → (e → TU a m)
          → TU a m
selectenv e s' s = (s e)
                   'choiceTU'
                   (s' e 'letTU' λe' →
                    oneTU (selectenv e' s' s))
```

The shown strategy combinator unites propagation with selection. It is an elaboration of the simpler traversal scheme *select*. It uses *choiceTU* and *oneTU* in the same manner as *select* (see Figure 4). All the additional behaviour directly implements the PROPAGATION pattern. The *selectenv* combinator is used in program analyses when a type-unifying node processor relies on environment propagation, e.g., to maintain bound variables along the way down to a focused fragment. For other predefined traversal schemes, a propagating version can be given in a similar way.

**Related Patterns**
Instead of using Propagation, you might use a reader monad to propagate information down the tree. The LOCAL EFFECT pattern explains how to keep the monadic propagation effect local.

## Generic Container — Advanced

**Intent**
Use a strategy as a generic data container.

**Motivation**
Sometimes terms of different types need to be stored in the same container. Such a generic container can be modelled with strategies.

**Schema**

```
type GC = ... S ...
emptyGC :: GC
emptyGC = ...
addGC :: Term t ⇒ t → GC → GC
addGC t c = ... modify c t ...
elemGC :: Term t ⇒ t → GC → Bool
elemGC t c = ... apply c t ...
...
```

**Description**
Just as monomorphic functions can be used to represent homogeneous data structures such as maps and sets, strategies can be used as heterogeneous data structures. Define your generic container type as a data structure that involves a strategy type $S$. Define the operations on your container in terms of strategy combinators. Operations modifying a container involve function modification lifted to the strategy level. Looking up data from a container involves strategy application.

**Sample code**
A generic container for assigning integer codes to terms.

```
type Coder = (Int, TU Int Maybe)
noCode :: Coder
noCode = (0, failTU)
getCode :: Term x ⇒ Coder → x → Maybe Int
getCode (_, s) = applyTU s
setCode :: (Term x, Eq x) ⇒ Coder → x → Int → Coder
setCode (i, s) x i' = (i, modifyTU s x (return i'))
nextCode :: Coder → (Int, Coder)
nextCode (i, s) = (i, (i + 1, s))
enCode :: (Term x, Eq x) ⇒ Coder → x → Coder
enCode c x = maybe gen found (getCode c x)
  where
    gen = let (i, c') = nextCode c in setCode c' x i
    found = const c
```

The type *Coder* assigns unique integers to terms of arbitrary types. It contains a counter as first component that records the highest code issued so far. The second component is a type-unifying strategy which represents the mapping from terms to codes assigned so far. We can provide an initial coder with *no* codes assigned, *get* a code of a term, *set* the code for a term, and generate the *next* code. When a given term is *enCode*n, the *Coder* is only modified if no code was previously assigned to the term.

**Consequences**
Generic containers can be made observable only per type, i.e., element retrieval or enumeration can only be done if the type of the elements are provided as input.

**Related Patterns**
Container operations are usually Keyhole Operations.

## Type Argument — Advanced

**Intent**
Parameterize behaviour by a type argument.

**Motivation**
Sometimes, you want your strategy to display type-specific behaviour even though it does not directly consume or produce any values that involve this specific type. By adding a type argument you can specify your type of choice.

**Schema**

```
type TypeArg a = ...
typeArg = ...

foo :: TypeArg a → S
foo ta = ... ta ...

fooT :: S
fooT = foo (typeArg :: TypeArg T)
```

**Description**
Type arguments have to be modelled as value arguments. The challenge is to prevent having to supply a value of the intended type. Type arguments can be modeled in various ways. In general, you need a dedicated type constructor *TypeArg*, and you need an actual representation of the type argument, say, *typeArg*. A strategy *foo* which is controlled by a type argument then takes an argument of type *TypeArg a*. The strategy employs the type argument *ta* to internally disambiguate unresolved polymorphism. An actual instance *fooT* of *foo* will simply construct the appropriate type argument by type annotation.

**Sample code**
A tick combinator for counting with a type argument

```
type TypeGuard a = a → ()
typeGuard = const ()
typeTick g
  = adhocTU (build 0) ((λ() → return 1) ∘ g)
```

Count subterms of type *HsDecl*

```
countHsDecls :: HsModule → Int
countHsDecls = runIdentity ∘ applyTU worker
  where
    worker = crush poly
    poly = typeTick (typeGuard :: TypeGuard HsDecl)
```

Here we model type arguments by functions with the intended type as domain, and *()* as co-domain. The *typeTick* combinator is controlled by a type argument which it uses internally to disambiguate the polymorphism of the non-generic argument of *adhocTU*. In fact, *typeTick* returns 1 if a term of the intended type is encountered, and 0 otherwise. The function *countHsDecls* counts all Haskell declarations within the given Haskell module. It is structured as a keyhole operation around an instantiation of *typeTick* that takes *HsDecl* as actual type argument. The type-unifying traversal scheme *crush* is used to traverse an input term with the instantiated type argument. Here, we rely on the fact that all numeric types (class *Num*) instantiate the *Monoid* class.

**Related Patterns**
Type arguments can be used for strategies that perform per-type element retrieval or enumeration on Generic Containers.

| META SCHEME | ADVANCED |
|---|---|

**Intent**
Parameterize an algorithm by higher-order strategies.

**Motivation**
The most basic and common way in which strategy combinators are parameterized is by plain strategies, i.e., constant combinators. This kind of parameterization is heavily used for traversal schemes to separate out type-specific behaviour. A more flexible algorithm can be obtained if a meta-scheme is established, i.e., when some parameters are non-constant combinators themselves. This allows you to vary, for instance the traversal scheme employed by an algorithm or the composition operators of its circuitry.

**Schema**

$$meta\ b_1\ \ldots\ b_n\ u_1\ \ldots\ u_m\ s_1\ \ldots\ s_k =$$
$$\ldots\ recurse\ \ldots$$
$$\textbf{where}$$
$$recurse = meta\ \ldots$$

**Description**
Parameterize your strategy definition by strategy combinators. There is potential for binary combinator arguments $b_1, \ldots, b_n$, unary combinator arguments $u_1, \ldots, u_m$, and plain strategy arguments $s_1, \ldots, s_k$. Combinators with more than two arguments are possible as well but note that all basic strategy combinators and most library schemes are unary or binary. You can introduce parameters for aspects such as traversal control, data-flow and control-flow. When you instantiate the higher-order parameters of a meta-scheme, you are turning it into a plain traversal scheme.

**Sample code**
A meta scheme for traversal with some instantiations

$$traverse\ o\ t\ s\ =\ s\ `o`\ t\ (traverse\ o\ t\ s)$$
$$totaltdS\ s\ =\ traverse\ bothS\ allS\ s$$
$$totalbuS\ s\ =\ traverse\ (flip\ bothS)\ allS\ s$$
$$oncetdS\ s\ =\ traverse\ choiceS\ oneS\ s$$
$$oncebuS\ s\ =\ traverse\ (flip\ choiceS)\ oneS\ s$$
$$stoptdS\ s\ =\ traverse\ choiceS\ allS\ s$$
$$stopbuS\ s\ =\ traverse\ (flip\ choiceS)\ allS\ s$$

The combinator *traverse* is a highly parameterized traversal scheme. It is parameterized in a binary combinator *o* for the composition of node processing and recursive descent. It is further parameterized in a unary combinator *t* to control the traversal in the sense of how to descend into subterms. Finally, *traverse* carries a nullary strategy argument *s* for node processing. Note that *traverse* is still neutral with respect to type-unification vs. type-preservation. In fact, the instantiations of *traverse* employ the overloaded basic strategy combinators according to Strafunski's *OverloadingTheme*, e.g., *allS* and *choiceS*. We use the postfix "...*S*" to point out overloading as opposed to commitment to either *TP* or *TU*. The first instantiation *totaltdS* can be resolved to an equivalent of either the type-preserving *topdown* or the type-unifying *crush* from Strafunski's *TraversalTheme* (see Figure 4).

## 4. CONCLUSION

*Contribution*

We have identified 6 basic and 7 advanced design patterns for generic programming with functional strategies. A programmer who knows just the basic ones will already to a large extent be able to take advantage of the Strafunski style of generic programming. These patterns are not far removed from the combinatorial styles of programming familiar to most functional programmers. Their added value is in the mixture of genericity and specificity, and in the support for generic traversal. By employing generic traversal, one can concisely deal with large syntaxes, formats and systems of datatypes. In addition to the basic ones, we have indicated an open-ended list of advanced design patterns that deal with sophisticated means of parameterization, composition and representation. We extracted the basic and advanced patterns from our applications in strategic programming, e.g., from those discussed in [17, 16, 15, 11]. These applications deal with program analyses and transformations for various languages such as Cobol, Haskell, and Java.

*Related work*

**Object-oriented design patterns** We have taken our inspiration from the literature on object-oriented design patterns [7]. As indicated in Section 3, we have made some modifications to accommodate the characteristics of functional programming. A general comparison reveals further differences. The object-oriented design patterns are predominantly concerned with code organization, distribution of responsibilities over classes and objects, tuning dependencies to maximize variability and maintainability. The functional design patterns are more concerned with issues of behaviour, parameterization, and reusability. We conjecture that these differences are (partly) due to the available abstraction mechanisms in both paradigms. As an aside, functional programming idioms have served as a source of inspiration for the formulation of some object-oriented design patterns [12]. The resulting patterns even have partly to do with generic traversal.

**Further functional design patterns** Our catalogue of patterns aims to communicate expertise in deploying the combinators of Strafunksi's strategy library. A range of combinator libraries are in existence, and these may profit from design patterns to give guidance to their users. These include libraries for parsing [9], pretty-printing [8], and polytypic programming [10]. For language embedding [20] and sorting morphisms [2] presentations have been given in a way that primary 'usage patterns', and recurring problems are discussed in some free format. One can recently observe an emerging interest to define patterns for functional programming (see the initiative [1]). In addition to the aforementioned domains, We envision that the following themes will definitely benefit from a design-pattern approach of explanation:

- programming with monad transformers,
- strictification,
- first-class polymorphism,
- dynamic typing,
- parallel and distributed programming.

*Future work*

**Functional program refactoring** Further inspiration can be taken from the object-oriented literature. The notion of *refactoring* [18, 6] seems particularly helpful. This is also proposed in [21]. In [13], the first author motivates and specifies a few functional program refactorings. The style and vocabulary employed in our functional pattern catalogue can serve as a starting point for the elaboration of a catalogue of refactorings for functional programs. In some pattern descriptions we have already hinted at how one design can be transformed into another, e.g., to extract a TRAVERSAL SCHEME from an application-specific TRAVERSAL. The formal foundations of functional program transformation are reasonably well-understood [3, 19] but a proper catalogue of refactorings is not available. This is true in particular for refactoring generic programs.

**Language processors as functional programs** Our impression is that functional strategies are very appropriate for language processing in general (i.e., refactoring tools, program optimizers, program analysers, metrics tools, etc.), and for *functional* language processing, in particular—as indicated by several Haskell examples in this paper, but see also [11]. The design patterns we presented should provide guidance to the functional language implementor and tool developer in applying strategies in these domains.

# APPENDIX

The following Haskell datatypes approximate the Haskell 98 abstract syntax as defined in the *hsparser* project (http://www.pms.informatik.uni-muenchen.de/mitarbeiter/panne/haskell_libs/hsparser.html). In the sample code in the paper, we refer to these datatypes. They are included here for easy reference.

Note that the mere size of this grammar, i.e., the number of types and data constructors, clearly demonstrates the benefits of *robustness* and *conciseness* of strategic programming. While our sample code includes non-trivial functionality for the entire Haskell syntax, only a handful of the types and data constructors needed to be mentioned explicitly.

Source locations (Line, Indentation)

   **data** $SrcLoc = SrcLoc\ Int\ Int$

Various kinds of names

   **newtype** $Module = Module\ String$
   **data** $HsQName = Qual\ Module\ HsName \mid UnQual\ HsName$
   **data** $HsName = HsIdent\ String \mid HsSymbol\ String \mid HsSpecial\ String$

Top-level structure of Haskell modules

   **data** $HsModule = HsModule\ Module\ (Maybe\ [HsExportSpec])$
                             $[HsImportDecl]\ [HsDecl]$
   **data** $HsExportSpec = HsEVar\ HsQName$
                        $\mid HsEAbs\ HsQName$
                        $\mid HsEThingAll\ HsQName$
                        $\mid HsEThingWith\ HsQName\ [HsQName]$
                        $\mid HsEModuleContents\ Module$
   **data** $HsImportDecl = HsImportDecl\ SrcLoc\ Module\ Bool$
                          $(Maybe\ Module)\ (Maybe\ (Bool, [HsImportSpec]))$
   **data** $HsImportSpec = HsIVar\ HsName$
                        $\mid HsIAbs\ HsName$
                        $\mid HsIThingAll\ HsName$
                        $\mid HsIThingWith\ HsName\ [HsName]$

All kinds of declarations

   **data** $HsDecl = HsTypeDecl\ SrcLoc\ HsName\ [HsName]\ HsType$
                 $\mid HsDataDecl\ SrcLoc\ HsContext\ HsName\ [HsName]$
                             $[HsConDecl]\ [HsQName]$
                 $\mid HsInfixDecl\ SrcLoc\ HsAssoc\ Int\ [HsName]$
                 $\mid HsNewTypeDecl\ SrcLoc\ HsContext\ HsName\ [HsName]$
                             $HsConDecl\ [HsQName]$
                 $\mid HsClassDecl\ SrcLoc\ HsQualType\ [HsDecl]$
                 $\mid HsInstDecl\ SrcLoc\ HsQualType\ [HsDecl]$
                 $\mid HsDefaultDecl\ SrcLoc\ HsType$
                 $\mid HsTypeSig\ SrcLoc\ [HsName]\ HsQualType$
                 $\mid HsFunBind\ SrcLoc\ [HsMatch]$
                 $\mid HsPatBind\ SrcLoc\ HsPat\ HsRhs\ [HsDecl]$
   **data** $HsConDecl = HsConDecl\ SrcLoc\ HsName\ [HsBangType]$
                     $\mid HsRecDecl\ SrcLoc\ HsName\ [([HsName], HsBangType)]$
   **data** $HsAssoc = HsAssocNone \mid HsAssocLeft \mid HsAssocRight$

Different layers of types

   **data** $HsBangType = HsBangedTy\ HsType \mid HsUnBangedTy\ HsType$
   **data** $HsQualType = HsQualType\ HsContext\ HsType \mid HsUnQualType\ HsType$
   **data** $HsType = HsTyFun\ HsType\ HsType$
                  $\mid HsTyTuple\ [HsType]$
                  $\mid HsTyApp\ HsType\ HsType$
                  $\mid HsTyVar\ HsName$
                  $\mid HsTyCon\ HsQName$
   **type** $HsContext = [HsAsst]$
   **type** $HsAsst = (HsQName, [HsType])$

Pattern-match cases

   **data** $HsMatch = HsMatch\ SrcLoc\ HsQName\ [HsPat]\ HsRhs\ [HsDecl]$
   **data** $HsRhs = HsUnGuardedRhs\ HsExp \mid HsGuardedRhss\ [HsGuardedRhs]$
   **data** $HsGuardedRhs = HsGuardedRhs\ SrcLoc\ HsExp\ HsExp$
   **data** $HsPat = HsPVar\ HsName$
                $\mid HsPLit\ HsLiteral$
                $\mid HsPNeg\ HsPat$
                $\mid HsPInfixApp\ HsPat\ HsQName\ HsPat$
                $\mid HsPApp\ HsQName\ [HsPat]$
                $\mid HsPTuple\ [HsPat]$
                $\mid HsPList\ [HsPat]$
                $\mid HsPParen\ HsPat$
                $\mid HsPRec\ HsQName\ [HsPatField]$
                $\mid HsPAsPat\ HsName\ HsPat$
                $\mid HsPWildCard$
                $\mid HsPIrrPat\ HsPat$
   **data** $HsPatField = HsPFieldPat\ HsQName\ HsPat$
   **data** $HsAlt = HsAlt\ SrcLoc\ HsPat\ HsGuardedAlts\ [HsDecl]$
   **data** $HsGuardedAlts = HsUnGuardedAlt\ HsExp$
                        $\mid HsGuardedAlts\ [HsGuardedAlt]$
   **data** $HsGuardedAlt = HsGuardedAlt\ SrcLoc\ HsExp\ HsExp$

All forms of expressions and literals

   **data** $HsExp = HsVar\ HsQName$
                $\mid HsCon\ HsQName$
                $\mid HsLit\ HsLiteral$
                $\mid HsInfixApp\ HsExp\ HsExp\ HsExp$
                $\mid HsApp\ HsExp\ HsExp$
                $\mid HsNegApp\ HsExp$
                $\mid HsLambda\ [HsPat]\ HsExp$
                $\mid HsLet\ [HsDecl]\ HsExp$
                $\mid HsIf\ HsExp\ HsExp\ HsExp$
                $\mid HsCase\ HsExp\ [HsAlt]$
                $\mid HsDo\ [HsStmt]$
                $\mid HsTuple\ [HsExp]$
                $\mid HsList\ [HsExp]$
                $\mid HsParen\ HsExp$
                $\mid HsLeftSection\ HsExp\ HsExp$
                $\mid HsRightSection\ HsExp\ HsExp$
                $\mid HsRecConstr\ HsQName\ [HsFieldUpdate]$
                $\mid HsRecUpdate\ HsExp\ [HsFieldUpdate]$
                $\mid HsEnumFrom\ HsExp$
                $\mid HsEnumFromTo\ HsExp\ HsExp$
                $\mid HsEnumFromThen\ HsExp\ HsExp$
                $\mid HsEnumFromThenTo\ HsExp\ HsExp\ HsExp$
                $\mid HsListComp\ HsExp\ [HsStmt]$
                $\mid HsExpTypeSig\ SrcLoc\ HsExp\ HsQualType$
                $\mid HsAsPat\ HsName\ HsExp$
                $\mid HsWildCard$
                $\mid HsIrrPat\ HsExp$
   **data** $HsStmt = HsGenerator\ HsPat\ HsExp$
                 $\mid HsQualifier\ HsExp$
                 $\mid HsLetStmt\ [HsDecl]$
   **data** $HsFieldUpdate = HsFieldUpdate\ HsQName\ HsExp$
   **data** $HsLiteral = HsInt\ Integer$
                     $\mid HsChar\ Char$
                     $\mid HsString\ String$
                     $\mid HsFrac\ Rational$